# Nuclear spin diffusion in the semiconductor TlTaS$_3$

A. M. Panich[1]*, C. L. Teske[2] and W. Bensch[2]

[1] Department of Physics, Ben-Gurion University of the Negev, P.O. Box 653, Beer Sheva 84105, Israel

[2] Institute of Inorganic Chemistry, Christian-Albrechts University, Kiel, Germany

* Corresponding author. Fax: +972-8-647203

e-mail: pan@bgumail.bgu.ac.il

**Abstract**

We report on a $^{203}$Tl and $^{205}$Tl nuclear magnetic resonance study of the chain ternary semiconductor TlTaS$_3$. We show that spin-lattice relaxation in this compound is driven by two contributions, namely by interactions of nuclear spins with thermally activated carriers and with localized electron spins. The latter mechanism dominates at lower temperature; at that, our measurements provide striking manifestation of the spin-diffusion-limited relaxation regime. The experimental data obtained allow us to estimate the spin diffusion coefficient.

## I. INTRODUCTION

Since in 1949 Bloembergen [1] suggested that the spin magnetization in a rigid lattice could be spatially transferred by means of the mutual flipping of neighboring spins due to dipole-dipole interaction terms of $I_i^+ I_j^-$ type, a number of insulating solids have been studied to test the validity of nuclear spin-lattice relaxation mediated by spin-diffusion (e.g., [2-7]). Nowadays, spin diffusion in semiconductors attracts considerable attention due to coming application of semiconductor structures in spintronics and quantum computing.

In the present paper, we report on an NMR study of nuclear spin diffusion in ternary compound $TlTaS_3$. This compound is of interest due to its low dimensionality, semiconducting and eventual photoconducting properties. Crystal structure of $TlTaS_3$ [8] is built from columns of edge-sharing $[Ta^{+5}S_4S_{2/2}]_2$ double octahedra running along [010] (Fig. 1). They are kept by the chains of $Tl^+$ ions that also run along the $b$ axis. Thallium atoms occupy only one crystallographic position. The compound is a semiconductor with an optical band gap $E_g = 0.78$ eV [8]. At the same time, electrical conductivity measurement [8] yields activation energy that is by the order of magnitude smaller than the band gap, indicating carrier excitation from shallow donor states. Our recent room temperature NMR study [9] of $TlTaS_3$ has shown noticeable indirect nuclear exchange coupling among thallium nuclei that is realized due to the overlap of the Tl electron wave functions across the intervening S atom. This result is in good agreement with the calculated electronic band structure [9]. In the present article, we report on the temperature dependent NMR measurements of $TlTaS_3$ in the external magnetic fields 1.227 and 8.0196 T. We measured NMR spectra and spin-lattice relaxation times of $^{203}Tl$ and $^{205}Tl$ isotopes, that have spin $I=½$ and are excellent probes in studying the electronic properties of compounds [10-15]. Our data show that the spin-lattice relaxation in TlTaS$_3$ is driven by a mutual play of two contributions. First of them is the interaction of nuclear spins with free carriers, thermally activated to the conduction band from localized electronic states in the forbidden gap. The second is represented by the nuclear spin coupling with localized electronic states and dominates at lower temperatures. At that, we reveal a striking manifestation of the spin-diffusion-limited relaxation regime and estimate the spin diffusion coefficient.

## II. EXPERIMENTAL

Sample preparation and characterization is described elsewhere [8]. $^{203}$Tl and $^{205}$Tl NMR measurements of a powdered TlTaS$_3$ sample were carried out in the temperature range from 70 to 290 K using Tecmag LIBRA and Tecmag APOLLO pulse solid state NMR spectrometers, a Varian electromagnet and an Oxford superconducting magnet. The spectra were measured in the external magnetic field $B_0 = 8.0196$ T using the spin echo method accompanied by phase cycling. Thallium NMR shifts are given relative to an aqueous 0.002 mol dm$^{-3}$ solution of TlNO$_3$, the position of which is assigned to the value of 0 ppm. In the magnetic field $B_0 = 8.0196$ T, it shows the $^{205}$Tl resonance at 196.9360 MHz. $^{203}$Tl and $^{205}$Tl spin-lattice relaxation times $T_1$ were measured in the external magnetic fields $B_0 = 1.227$ and 8.0196 T by means of a saturation comb pulse sequence. The duration of the $\pi/2$ pulse was 2 μs. In order to measure the anisotropy of the spin-lattice relaxation rate in $B_0 = 8.0196$ T, we set the spectrometer frequency in series at the center and shoulders of the spectrum and used soft $\pi/2$ pulses of 33 μs at reduced radio frequency power.

## 3. RESULTS OF THE EXPERIMENTS AND DISCUSSION

$^{205}$Tl NMR spectra in $B_0 = 8.0196$ T (Fig. 2) correspond to a non-axially symmetric shielding tensor. At T = 290 K, its principal values are $\sigma_{11} = 640$ ppm, $\sigma_{22} = 1073$ ppm and $\sigma_{33} = 1499$ ppm (the precision is about 10%), the center of gravity is located at $\sigma_{iso} = 1071$ ppm, while at T = 70.5 K the principal values are $\sigma_{11} = 424$ ppm, $\sigma_{22} = 945$ ppm, $\sigma_{33} = 1407$ ppm, and $\sigma_{iso} = 925$ ppm. The asymmetry parameter is $\eta \sim 1$. In low magnetic field the contribution of chemical shielding anisotropy is small, and the line shape is determined by indirect exchange coupling among the spins of Tl nuclei [9].

Temperature dependences of the $^{205}$Tl spin-lattice relaxation time $T_1$ in the external magnetic fields $B_0 = 8.0196$ and $1.227$ T are shown in Fig. 3. In high magnetic field, $T_1$ exhibits some anisotropy resulting from the anisotropy of the electronic structure of the compound. The relaxation data obtained show several features. First, usually observed room temperature values of $T_1(^{205}$Tl$)$ in thallium-based semiconductors [15,16] vary from several seconds to several tens of seconds. However, one can find from Fig. 3 that the compound under study shows much shorter (by several orders of magnitude) values of $T_1$. Second, the temperature dependence of the relaxation time $T_1$ in the range from 155 to 300 K exhibits activation behavior, which is evidently caused by the increase in the number of carriers due to their thermal excitation to the conduction band. At that, the value of the activation energy, calculated from the slope of the $\log T_1(1/T)$ curve, was found to be $0.06 \pm 0.006$ eV, which is much smaller than the band gap. These two facts, the strongly enhanced nuclear spin-lattice relaxation and small value of the activation energy, show that the observed relaxation behavior should be assigned to the interaction of nuclear spins with free carriers thermally activated to the conduction band from some localized electronic states in the forbidden gap, e.g. shallow donor states located below the bottom of the conduction band. This conclusion is in good agreement with the conductivity measurements [8].

Furthermore, comparing the temperature dependences of $^{205}$Tl spin-lattice relaxation time in the external magnetic fields $B_0 = 8.0196$ and $1.227$ T (Fig. 3), one can find that $T_1$ increases with increasing magnetic field, while the theory of spin-lattice relaxation via conduction electrons [17] predicts field-independent $T_1$. It means that besides the relaxation via conduction electrons, another mechanism takes place. Among the relaxation mechanisms relevant to the compound under study, a characteristic feature of the Raman process is independence of the relaxation time on the applied magnetic field, while a direct relaxation process usually predicts very long relaxation times ($> 10^4$ s) and yields reduction in $T_1$ with increasing magnetic field as $T_1 \sim B_0^{-2}$ (Ref. 17). Analogous $T_1 \sim B_0^{-2}$ dependence [17] is characteristic of the contribution of chemical shielding anisotropy. The only mechanism that causes increase in $T_1$ with increasing magnetic field is the heat contact with the lattice via localized electronic states (e.g., paramagnetic centers). This mechanism yields either $T_1 \sim B_0^2$ or $T_1 \sim B_0^{1/2}$ in the cases of rapid spin diffusion and diffusion-limited relaxation, respectively [4]. Therefore one is led to conclusion that along with the spin-lattice relaxation via thermally activated carriers, the relaxation via the localized electronic states has also to be taken into account. The latter contribution should be stronger at low temperature when electrons are localized, likely behaving as paramagnetic centers, and weaker at high temperature when electrons are excited to the conduction band. It is readily confirmed by the experimental fact that the $T_1$ values, measured in different magnetic fields (Fig. 3), are close to each other at higher temperatures, where the relaxation by means of thermally activated carriers dominates, but become significantly different at lower temperatures, where the relaxation by means of localized electronic states dominates.

The latter relaxation mechanism is known to be mediated by spin diffusion that occurs through the mutual flips between neighbouring nuclear spins and results in the magnetization transfer from the distant nuclear spins to the localized electron spins. The relaxation that involves nuclear spin diffusion is notoriously difficult to pin down experimentally. However, in our experiments, convincing evidence of the connection between spin diffusion and spin-lattice relaxation is provided by a comparative study of the relaxation times of two thallium isotopes measured in the same magnetic field (Fig. 4). One can find from Fig. 4 that $T_1(^{203}Tl) > T_1(^{205}Tl)$, and the distinction between $T_1$'s becomes significant below 150 K. At that, we note that both Tl isotopes have spin I=½, and their gyromagnetic ratios differ only by one percent ($\gamma(^{205}Tl)/\gamma(^{203}Tl)=1.01$), that evidently cannot cause such noticeable difference in $T_1$'s. The only explanation of this effect is that the observed difference comes from a relaxation mechanism that is mediated by spin diffusion, a process that requires two adjacent *like* nuclei to exchange their spin quantum numbers. Indeed, the natural abundances of $^{203}Tl$ and $^{205}Tl$ isotopes are 29.5 and 70.5%, respectively. The average distance between the $^{205}Tl$ nuclei is therefore smaller than that between the $^{203}Tl$ nuclei. The probability of the mutual spin flip decreases as the distance between the interacting nuclei increases. Therefore the spin-diffusion-assisted relaxation of the $^{203}Tl$ nuclei should be less effective, and their $T_1$ should be longer in comparison with that of $^{205}Tl$ nuclei; as well, the diffusion coefficient $D(^{203}Tl)$ should be smaller than $D(^{205}Tl)$. This is observed in our experiment (Fig. 4), providing a striking manifestation of the spin-diffusion-limited relaxation regime.

Usually in the literature spin diffusion is considered as driven by the dipole-dipole interaction that causes the mutual flips of adjacent nuclear spins. However, for thallium atoms, the indirect nuclear exchange coupling is the dominating line-broadening mechanism [9-15]. Since the exchange interaction Hamiltonian

$$H = \sum_{i,j} J_{ij}[I_i^z I_j^z + \frac{1}{2}(I_i^+ I_j^- + I_i^- I_j^+)] \qquad (1)$$

also contains the mutual spin-flip operator $I_i^+ I_j^-$, spin diffusion in the case in question is evidently mediated by the indirect exchange coupling among Tl nuclei. Taking into account the scalar term of the exchange Hamiltonian (Eq.1) only and using the procedure analogous to that of Ref. 3, one can derive an expression for the diffusion coefficient

$$D = \frac{a^2 J^2 T_2}{4\sqrt{2}} \qquad (2)$$

where $a$ is the distance between adjacent nuclei, J is the exchange coupling parameter, and $T_2$ is the spin-spin relaxation time. Using the structural data of $TlTaS_3$ [8], J(Tl-Tl)=8.6 kHz [9], and $T_2(^{205}Tl)$ = 62 μs at 90 K, we estimated the diffusion coefficient as D=2.1 ×10$^{-12}$ cm$^2$/s.

Let us now discuss the magnetization recovery plots of two thallium isotopes (Fig. 5). According to the theory of the spin-diffusion-limited relaxation, this process comprises two regimes. Since initially the magnetization is saturated and there is no gradient of magnetization density, at the beginning of the relaxation process spin diffusion is not of importance, and thus the direct relaxation dominates causing stretch exponential time dependence. Then this curve proceeds asymptotically to an exponential function of time. This is the region where spin diffusion dominates and results in equilibrium spin magnetization. These two regimes are seen in

Fig. 5. They join at the time $t_b = C^{1/2}D^{-3/2}$ (Ref. 2), corresponding to a distance $b=(C/D)^{1/4}$ from the localized electron spin. One can find from Fig. 5 that the values of $t_b$ are different for two thallium isotopes. This is evidently due to the difference in their diffusion coefficients. Experiment yields $t_b(^{203}\text{Tl})/t_b(^{203}\text{Tl}) \approx 3.0 \pm 0.4$, thus $D(^{205}\text{Tl})/D(^{203}\text{Tl}) \approx 2.1$.

## 4. SUMMARY

In summary, our $^{203}$Tl and $^{205}$Tl NMR study of the ternary semiconductor TlTaS$_3$ shows that the spin-lattice relaxation time $T_1$ is driven by two contributions, namely by the interactions of nuclear spins with the thermally activated carriers and with localized electronic spins. The former mechanism dominates at higher temperatures, when electrons are excited to the conduction band, while the latter dominates at lower temperatures, when electrons are localized. At that, the spin-lattice relaxation rate is limited by spin diffusion caused by the indirect exchange coupling of nuclear spins. A striking manifestation of the spin-diffusion-limited relaxation regime becomes apparent from the comparative study of $T_1$'s of two thallium isotopes measured in the same magnetic field. The formula for spin diffusion coefficient $D$ in the case of exchange coupling is derived, and the value of $D$ is estimated.

**Acknowledgment**

One of the authors (A.M.P.) thanks A. Pines, G. B. Furman and V. A. Atsarkin for helpful discussions and comments.

**Figure legends.**

Fig. 1. Crystal structure of TlTaS$_3$ (perspective view along [010]). Some of the Tl-S bonds are shown by dotted lines.

Fig. 2. $^{205}$Tl NMR spectra of the powder TlTaS$_3$ sample in magnetic field B$_0$=8.0196 T at different temperatures.

Fig. 3. Semi-logarithmic plots of the $^{205}$Tl NMR spin-lattice relaxation time *versus* reciprocal temperature in powder TlTaS$_3$ in the magnetic fields B$_0$=8.0196 and 1.227 T. Squares, circles and triangles correspond to the low, medium and high frequency singularities of the $^{205}$Tl NMR spectrum in B$_0$=8.0196 T.

Fig. 4. Semi-logarithmic plots of the $^{203}$Tl and $^{205}$Tl NMR spin-lattice relaxation time *versus* reciprocal temperature in the magnetic field 1.227 T.

Fig. 5. Semi-logarithmic plots of the magnetization recoveries of the $^{203}$Tl and $^{205}$Tl isotopes.

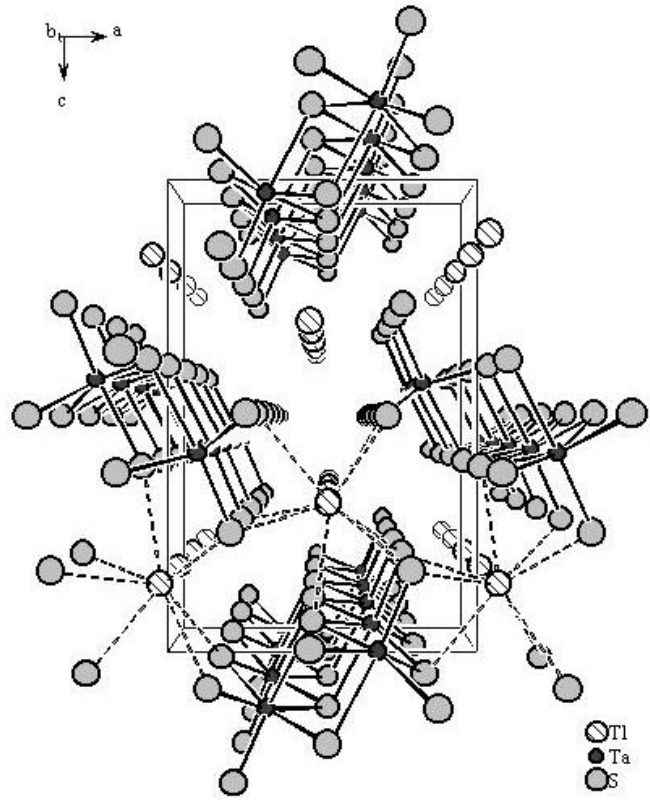

Fig. 1.

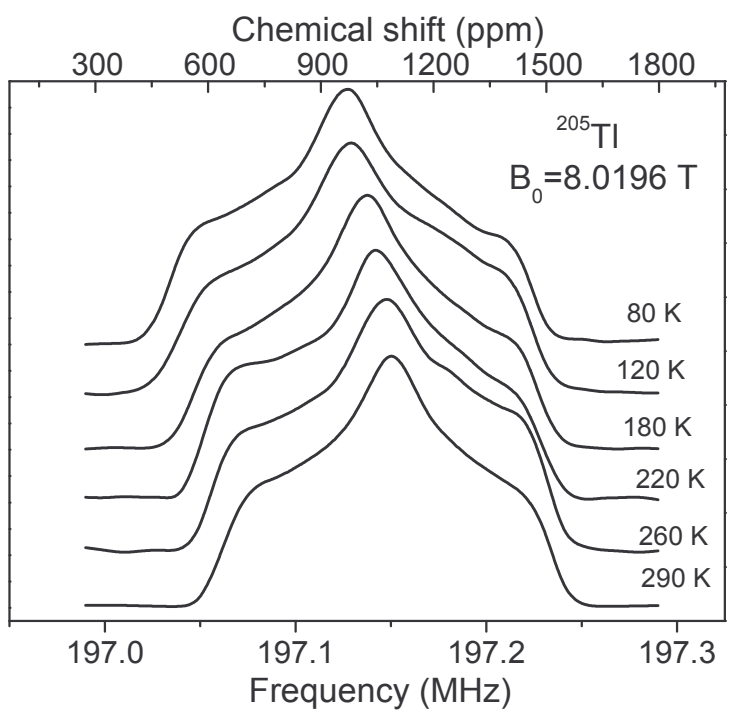

Fig. 2.

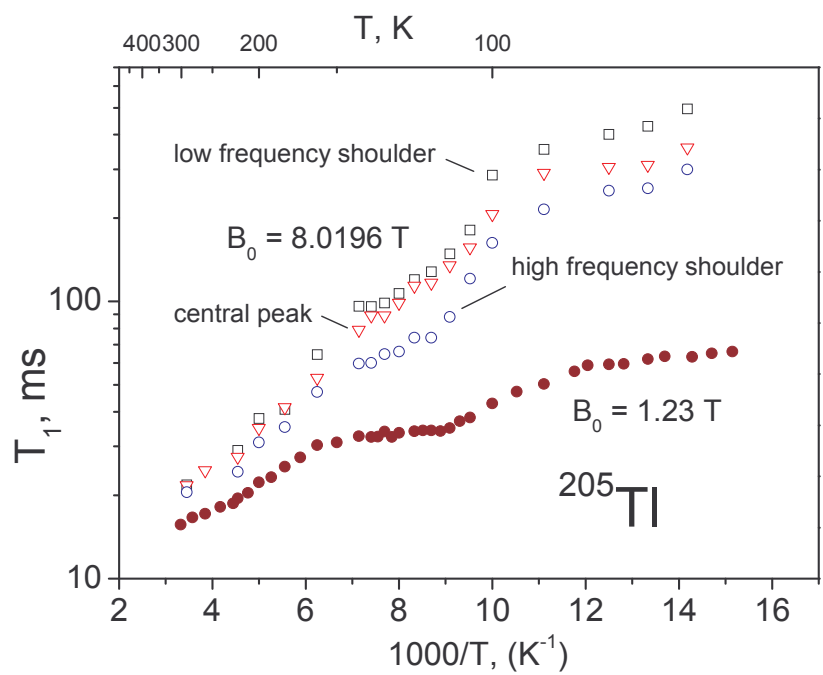

Fig. 3.

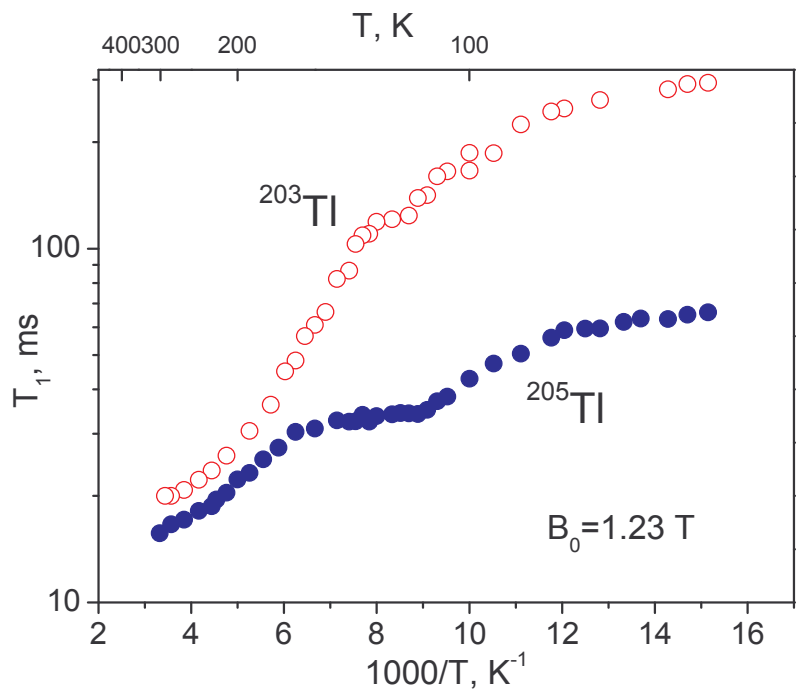

Fig. 4.

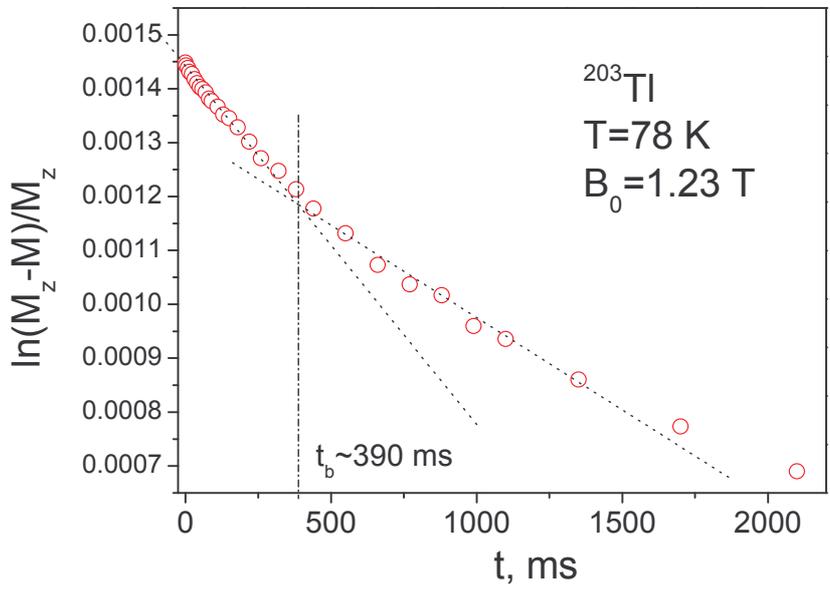

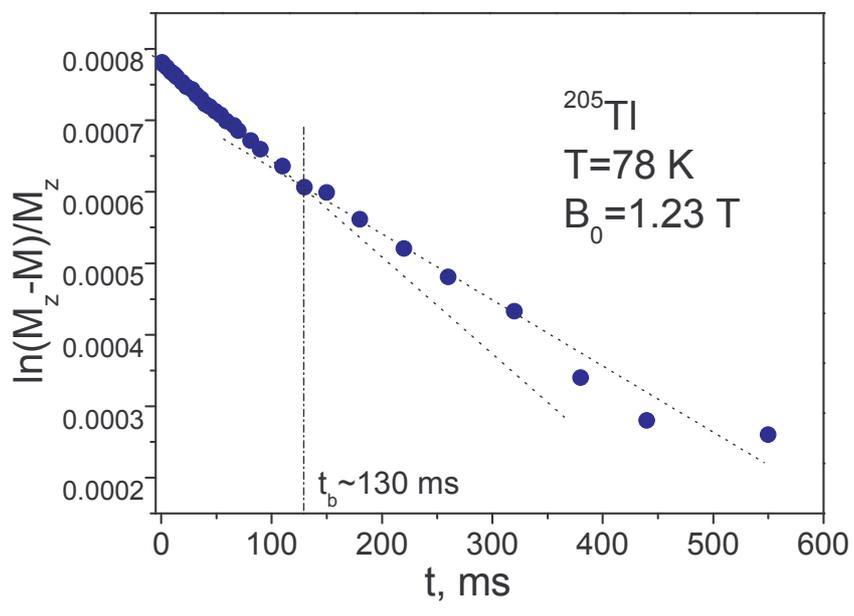

Fig. 5.